# Reliability of Redundant M-Out-Of-N Architectures With Dependent Components: A Comprehensible Approach With Monte Carlo Simulation

Tim Maurice Julitz[a], Antoine Tordeux[b], Nadine Schlüter[a], Manuel Löwer[a]

[a]*Department of Product Safety and Quality Engineering, University of Wuppertal, Germany*
[b]*Department of Traffic Safety and Reliability, University of Wuppertal, Germany*

**Abstract**

Redundant architectures can improve the reliability of complex systems. However, component dependencies can affect the architecture and negate the benefit of redundancy. In this paper, we develop three component dependency models and analyze the reliability of different M-out-of-N configurations using Monte Carlo simulation. The first model assumes a linear component dependency. The second and third models consider common cause failures, in the latter for all components and in the second for random groups of components. As expected, the results show that interdependency degrades the reliability of parallel 1ooN systems while improving it for serial NooN systems. Interestingly, 2oo3 systems produce intermediate results that show an improvement in reliability for certain indicators and a deterioration for some others, depending on the type of dependency models. The results show nonlinear properties of MooN systems with dependent components, which suggest careful handling in applications. An online simulation platform based on Monte Carlo Simulation enables product designers to use the models efficiently and achieve tailored results.

## 1. Introduction

### 1.1. Dependency in Redundant System Architectures

Redundancy is a common approach to increase the reliability of systems. Consequently, redundant system architectures are becoming increasingly necessary in modern products with a growing demand for safety. Redundancy can be implemented at the component level (serial-parallel systems) or at the system level (parallel-serial-systems). The Barlow-Proschan (BP) principle states that redundancy at the component level is generally more reliable than redundancy at the system level (Barlow and Proschan, 1981). These redundant configurations are discussed extensively in the literature. Safety-critical industries like aerospace, rail transportation, nuclear energy and the automotive sector use M-out-of-N architectures (also called MooN, K-out-of-N, voting or majority redundancy). Their field of application is broad and therefore highly relevant. MooN systems can be used in domains where fail-operational functions are required. They enable self-diagnosis of systems by comparing redundant inputs according to their logic. A MooN system consisting of *N* components is operational if at least M components are working. Therefore, MooN architectures are particularly relevant for the area of autonomous driving (Julitz et al., 2023). It is frequently assumed that the system's components are independent, but this assumption does not apply in reality. Components of real systems are interdependent, e.g. as they are exposed to the same environment or share the same load. The functional safety standard of the automotive industry requires the consideration of dependencies caused by shared inputs, communication, interface, shared resources, components of identical type, systematic coupling and insufficient environmental immunity (ISO, 2018). Preliminary results for multisensory perception in autonomous vehicles have shown that component dependence can degrade the benefits of redundant architectures in terms of reliability, e.g. in situations where common cause failures can occur (Gottschalk et al., 2022). In other situations, the reliability can be significantly improved, e.g. when different sensors complement each other (Kowol et al., 2021). In general, the impact of component dependency of redundant architectures on system reliability is still poorly understood in reliability engineering, especially with regard to MooN architectures (Fang and Li, 2016). Product designers must consider the possible dependencies between components. Existing dependency models (Section 1.2) are complex and difficult to understand. User-oriented tools are required for comprehensible use in product development.

**1.2. Related Work: Models of Dependent Failure Behavior**

Several models of dependent failure behaviour have been proposed. A literature review has identified that they fall into five categories: Lifetime distribution models, system state models and degradation process models (statistical association among the variables) as well as failure interaction models and failure propagation models (mechanistic dependency models) (Zeng et al., 2023).

In **Lifetime distribution models**, distributions are joint to model dependency. (Kelkinnama and Asadi, 2022) evaluated the reliability of a system consisting of n components grouped into L different types. Each group has its individual distribution, which is linked via a copula function. (Fang and Li, 2017) analysed component lifetimes $X$ with matched redundancy component lifetimes $Y$ which are distributed differently and linked with a copula factor. (Gupta and Kumar, 2014) modelled dependency by distributing components and their spare parts differently considering failure rate, likelihood and stochastic ordering. (Jeddi and Doostparast, 2016) used joint distributions of component lifetimes to analyse two dependent components of serial and parallel systems. (Kotz et al., 2003) investigated the influence of the degree of correlation of dependent components on parallel system lifetimes. This is done by comparing bivariate distributions. (Navarro and Fernández-Martínez, 2021) represented dependency through copula and distortion functions which couples a set of functions of individual distributions. The components of each distribution function are dependent. The copula is based on the dependence structure. (Yan et al., 2022) studied serial-parallel and parallel-serial systems and their relationship with a copula coefficient which couples the distribution of original components with the distributions of their spare parts. (Zhang et al., 2017) compares distributions of the same and different types with a copula scalar factor in a multivariate distribution and confirms the BP-principle. The paper considers hazard rates, likelihoods and ratio orderings.

**System state models** focus on representing states of components or systems. (Agnieszka and Bozena, 2016) modelled a multistate parallel-serial system with dependent components. The states represent characteristic reliability functions which are combined in a multistate reliability function. The dependency of spare parts on original components is implemented with a stress proportionality correction coefficient. (Li et al., 2022) uses state-dependent lifetime distributions. The states are modelled with co-variates derived from environmental conditions such as frequency, temperature or weather. (Xing and Levitin, 2013) modelled system states (phases) with fault trees to identify common cause failures. (Mi et al., 2018) analysed the reliability of multi-state systems with Bayesian Networks, in which the nodes represent systems states. Dependent failures are determined by the fraction of the total failure probability attributable to dependent failures.

**Degradation process models** refer to processes in which the reliability of a system or component decreases over time. Statistical processes like Wiener, gamma or inverse Gaussian can be used for modelling degradation processes (Zeng et al., 2023). Also, the combination of random shock processes and degradation processes or applications of copula models for degradation processes can be found in the literature (Zeng et al., 2023).

**Failure interaction models** explain how different failure mechanisms interact with each other. (Fang and Li, 2016) focus on the system structure of serial and parallel systems and their structural dependencies which lead to failures. A Bayesian network approach by (Cai et al., 2012) evaluates common cause failures that could lead to failures in a MooN system using binary tables of component failures.

**Failure propagation models** are used to describe a chain of events or processes that lead to a failure, which is also called cascading failure and is a subset of common cause failures. (Da Costa Bueno and Martins do Carmo, 2007) modelled failure propagation of a MooN system with a compensator process. Based on known component lifetimes, conditional probabilities of failures are projected. Self-evolution processes, network models and simulations are also used to model failure propagation (Zeng et al., 2023).

A literature review by (Dhillon and Anude, 1994) has compiled **modeling techniques for common cause failures** (CCF). The β-factor method, the binomial failure rate method, the square root bounding method, system-specific methods and fault tree-based approaches with cut-set analysis are discussed. Markov models are highlighted for stochastic analysis of CCF. Challenges for the practical identification of common cause failures include the number of possible causes of CCF that need to be identified, the suitability of modeling techniques for reliability analysis, available data of CCF events and a consistent definition of CCFs (Dhillon and Anude, 1994).

### 1.3. Research Objective

A variety of approaches have been explored to address the issue of dependent failures. Typically, they are restricted to series or parallel systems, or limited to specific probability distributions. This makes it difficult to find generic properties, resulting in specific solutions for each model. A comprehensive understanding of how component dependencies affect the reliability of the entire system remains elusive. A research gap emerges especially when the impact of dependencies in MooN architectures are considered. Most of the preliminary work is based on complex and complicated mathematical models. Their efficient application can pose a challenge for the product designer. Therefore, an efficient model for studying component dependencies should be developed with a focus on user-friendliness. The model is intended to contribute to a better understanding of dependencies in redundant architectures, also with regard to MooN systems, and to provide general insights for cross-domain applications. To this end, the following questions are examined:

- How can the dependent failure behaviour of systems be modelled in a minimal way?
- What generalizations can be made about the impact of dependent redundancy of serial, parallel, and M-out-of-N architectures on system reliability?
- How can the model be used in a user-friendly way?

The paper is structured as follows. In Section 2, component dependency models are developed in terms of pairwise linear dependency, global common cause failures where all components fail, and marginal common cause failures where some components fail. A copula factor $X_0$ is used. Section 3 presents the results for a serial, parallel and a MooN system. Section 4 describes how the results are obtained und how the models can be used. For this purpose, a script is written in R (Figure 7), which is integrated into an online simulation platform. A conclusion is given in Section 5.

## 2. Component Dependency Models

In the following, three linear models of component interdependence including a covariate are developed and analysed for multivariate systems with $N > 1$ components. The dependent times to failure of the components are given by Equation (1).

$$Y = (Y_k, k = 1, \ldots, N). \tag{1}$$

They are determined on the basis of $N + 1$ independent and identically distributed continuous random variables (typically Weibull distributed) given by Equation (2).

$$X = (X_k, k = 0, \ldots, N). \tag{2}$$

The dependency is quantified by a parameter $p \in [0,1]$, the component times to failure being independent if $p = 0$ and equal if $p = 1$. The time to failure T of a $MooN$ ($M$-out-of-$N$) system, which remains operational until at least $M$ components are operational, is the $M/N$ quantile of the vector $Y$ as given by Equation (3).

$$T = q_{M/N}(Y). \tag{3}$$

### 2.1. Linear Dependency Model

First, a linear independence of the component times to failure is assumed, including a covariate $X_0$, weighted by the parameter $p \in [0,1]$, see Equation (4).

$$Y_k = (1 - p)X_k + pX_0, \quad k = 1, \ldots, N. \tag{4}$$

The component times to failure $Y_k = X_k$, $k = 1,\ldots,N$, are independent if $p = 0$ while they are equal to $Y_k = X_0$ for all $k = 1,\ldots,N$ if $p = 1$. Here, except for the later extreme case where $p = 1$, the component times to failure, being random continuous, cannot be strictly equal.

## 2.2. Global Common Cause Failure Model

In contrast to the previous linear model, the component interdependency is modelled through global common cause failure possibilities for which all components can fail simultaneously. This corresponds to cases of damage affecting the entire system or subsystems. as shown in Equation (5), it is no longer a question of a linear combination of variables but of a probability $p$ that all the components fail at the same time $X_0$.

$$Y_k = (1 - \xi)X_k + \xi X_0, \quad \xi \sim \mathcal{B}(p), \quad k = 1,\ldots,N. \tag{5}$$

Where $\xi$ is an independent Bernoulli random variable equal to zero with probability $p$ and to one with probability $1 - p$. Two cases are possible with probability $p$ in this modelling framework: either all the component times to failure $Y$ are equal to $X_0$, or they are all independent and equal to $X$.

## 2.3. Marginal Common Cause Failure Model

All components can fail simultaneously with the model in Equation (5). Alternatively, it can also be assumed that only some of the components fail simultaneously. In this case, the common cause failure probability is specific to each component as described by Equation (6).

$$Y_k = (1 - \xi_k)X_k + \xi_k X_0, \quad \xi_k \sim \mathcal{B}(p), \quad k = 1,\ldots,N. \tag{6}$$

As before, $\xi_k$, $k = 1,\ldots,N$, are independent Bernoulli random variables with the parameter $p$. In this model, the component times to failure are equal to the covariate $X_0$ with probability $p$ or are independent and equal to the random variables $X$ with probability $1 - p$.

## 3. Results

In the following, a Monte Carlo simulation is used to analyse the distribution of the MooN system time to failure $T$ with $N = 3$ components for the three interdependency models given in Equations (4), (5), and (6). The analysis is performed for $p$ ranging from zero to one, so that the system is an ordinary MooN system with independent components for $p = 0$, while it resumes to a system with a single component for $p = 1$. A systematic methodological procedure is followed: First, the probability density function (PDF) of the system times to failure and the system reliability is drawn. The mean, median, mode and standard deviation statistics are the presented in relation to the performance of an ordinary MooN system with independent components for which $p = 0$. This allows quantifying the proportional impact of component interdependency on the reliability and time to failure of MooN system. Three systems with $N = 3$ components are considered:

- The case $M = 1$ corresponds to a 1oo3 parallel system that fails when all the three components are no longer operational.
- The case $M = 2$ corresponds to a 2oo3 system, also referred to in the literature as a 1-ECU fail-operational architecture.
- The case $M = 3$ corresponds to a 3oo3 series system that fails when one of the three components fails.

## 3.1. Linear Dependency Model

It is noteworthy that, since the expected value is a linear operator, the model linearity is preserved on average with the linear dependency model Equation (4). In fact, as shown in Equation (7), using this model results in

$$E(T) = E(Q_{M/N}(Y)) = E(Q_{M/N}((1-p)X_k + pX_0))$$
$$= (1-p)E(Q_{M/N}(X_k)) + pE(X_0). \tag{7}$$

The remaining statistics, namely median, mode, and standard deviation, are nonlinear operators. Therefore, nonlinear effects should occur by varying the dependency parameter $p$. Figure 1 shows the probability density function (PDF) and the reliability function while Figure 2 shows the location statistics for the time to failure according to the dependency parameter $p$.

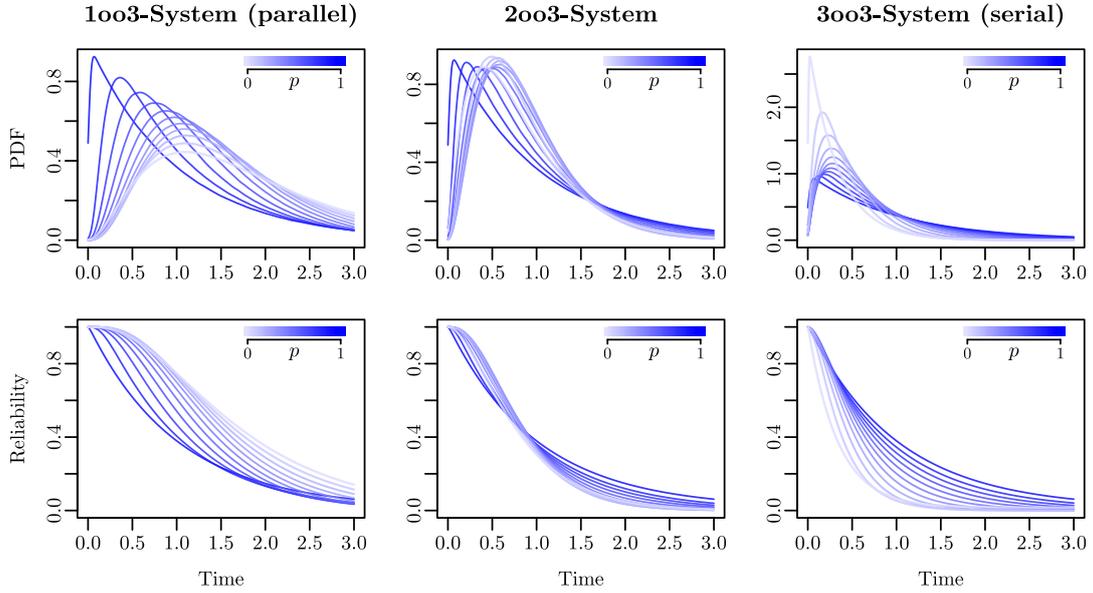

Fig. 1. Probability density function (PDF) of the time to failure (top panels) and reliability function (bottom panels) for the M-out-of-N systems with the linear dependency model Equation (4) and $p$ ranging from 0 to 1.

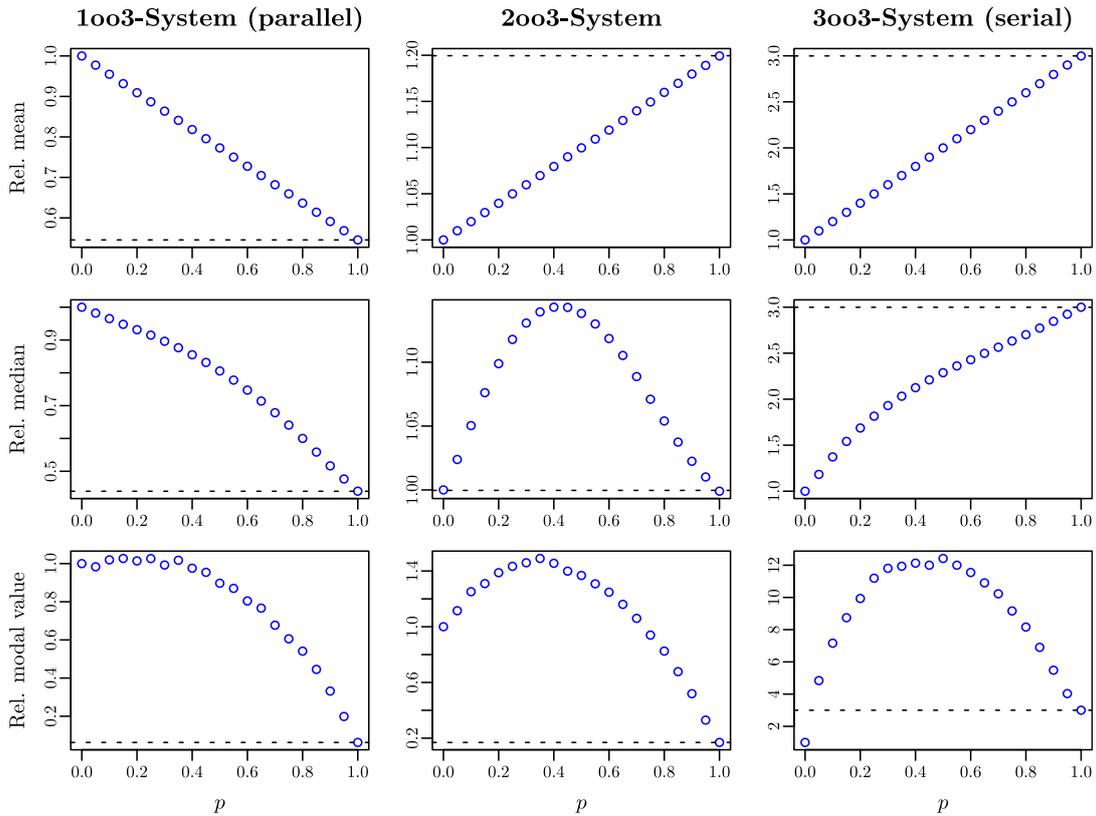

Fig. 2. Relative mean value, median and mode (from the top to the bottom) of the time to failure of M-out-of-N systems with the linear model Equation (4) and $p$ ranging from 0 to 1.

The dependency systematically degrades the reliability and related statistics (mean, median and modal time to failure) for parallel system 1oo3 (Figures 1 and 2, left panels), while it systematically improves (up to the modal value) for the serial system 3oo3 (Figures 1 and 2, right panels). The 2oo3 system shows mitigated results: improvement of the mean time to failure but deterioration of the modal value and nonlinear unimodal relationship for the median (Figures 1 and 2, middle panels). As expected, the mean values show a linear behaviour, when the mode and median are nonlinear for all the systems (see Figure 2).

### 3.2. Global Common Cause Failure Model

As with the previous linear model, the global common cause failure interdependence model equation (5) preserves linearity over $p$ on average. Indeed, as shown in Equation (8), using this model results in

$$E(T) = E(Q_{M/N}(Y)) = E(Q_{M/N}((1-\xi)X_k + \xi X_0))$$
$$= (1-p)E(Q_{M/N}(X_k)) + pE(X_0), \tag{8}$$

since $E(\xi) = p$. As previously, the remaining statistics such as median, modal value and standard deviation, which are nonlinear operators, have no obvious linear relationship with $p$. Figure 3 shows the PDF and the reliability function of the CCF dependence models, while the time to failure location statistics are shown in Figure 4.

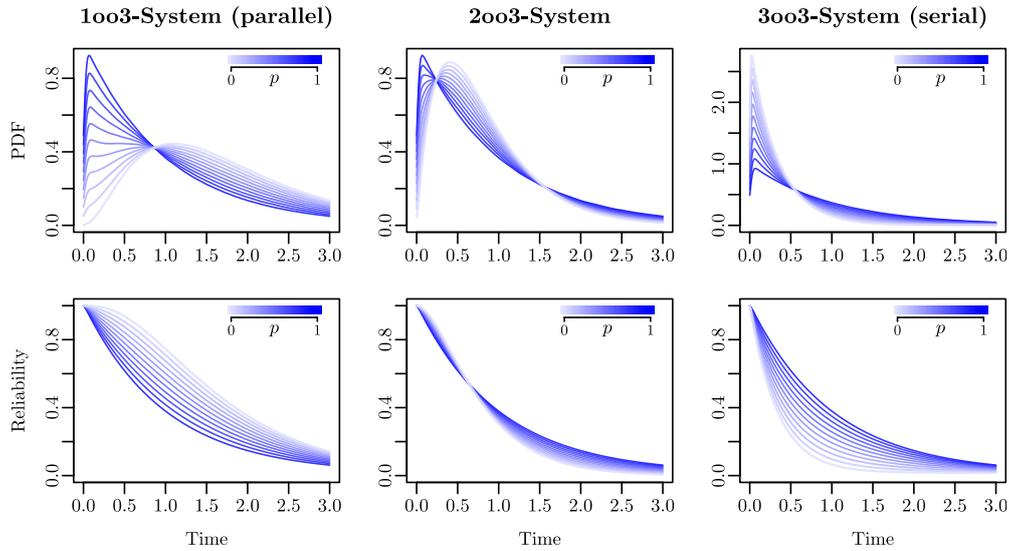

Fig. 3. Probability density function (PDF) of the time to failure (top panels) and reliability function (bottom panels) for the M-out-of-N systems with the global Common Cause Failure Model Equation (5) and $p$ ranging from 0 to 1.

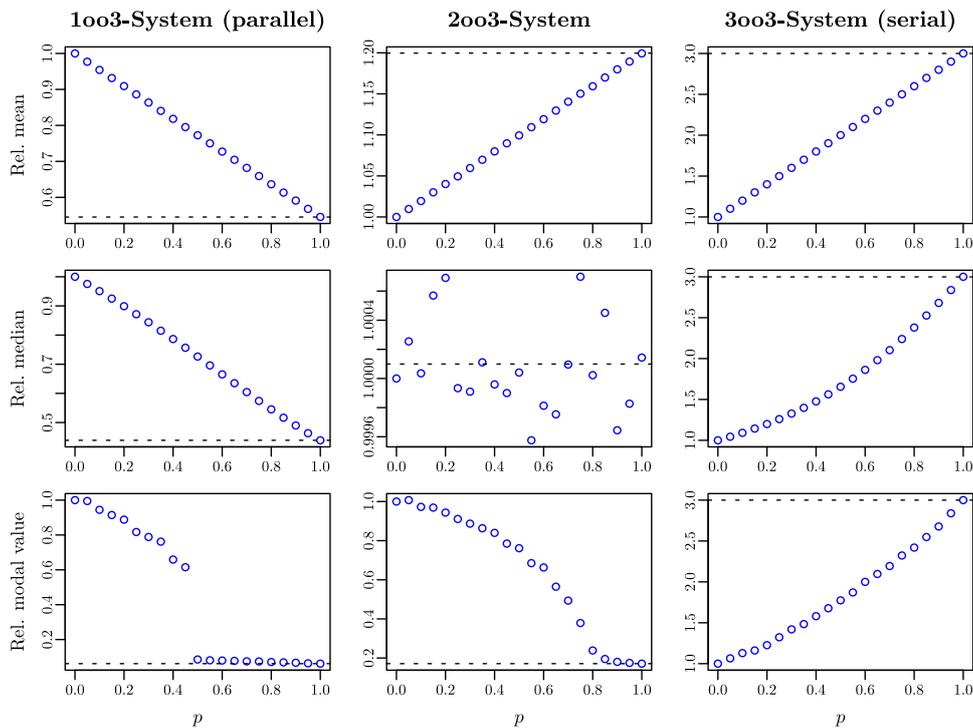

Fig. 4. Probability density function (PDF) of the time to failure (top panels) and reliability function (bottom panels) for the M-out-of-N systems with the global Common Cause Failure Model Equation (5) and p ranging from 0 to 1.

As for the linear model equation (4), the dependency deteriorates the reliability of the 1oo3 parallel system while it improves it for the 3oo3 serial system degradation (Figures 3 and 4, right panels). The 2oo3 redundant system shows intermediate results (Figures 3 and 4, middle panels). More precisely, a nonlinear transition can be observed as $p$ increases, from unimodal distributions with large location and low variability to asymmetric distributions with low mode and large right tails for the redundant 1oo3 und 2oo3 systems. Note that the median is invariant with $p$ for the 2oo3 system.

### 3.3. Marginal common cause failure model

In contrast to the previous models, the marginal common cause failure dependency model equation (6) no longer preserves linearity over $p$ on average. In fact, this results in Equation (9),

$$\begin{aligned} E(T) = E(Q_{M/N}(Y)) &= E(Q_{M/N}((1-\xi_k)X_k + \xi X_0)) \\ &\neq (1-p)E(Q_{M/N}(X_k)) + pE(X_0), \end{aligned} \quad (9)$$

as $\xi_k$ is specific to the component $k$.

Qualitatively, the marginal CCF model shows similar behaviors to the linear and global CCF models (see Figures 5 and 6). In particular, the 2oo3 system still presents mitigated results, with improvements in the modal time to failure, deterioration of the mean and no effect on the median (Figure 6, middle panels).

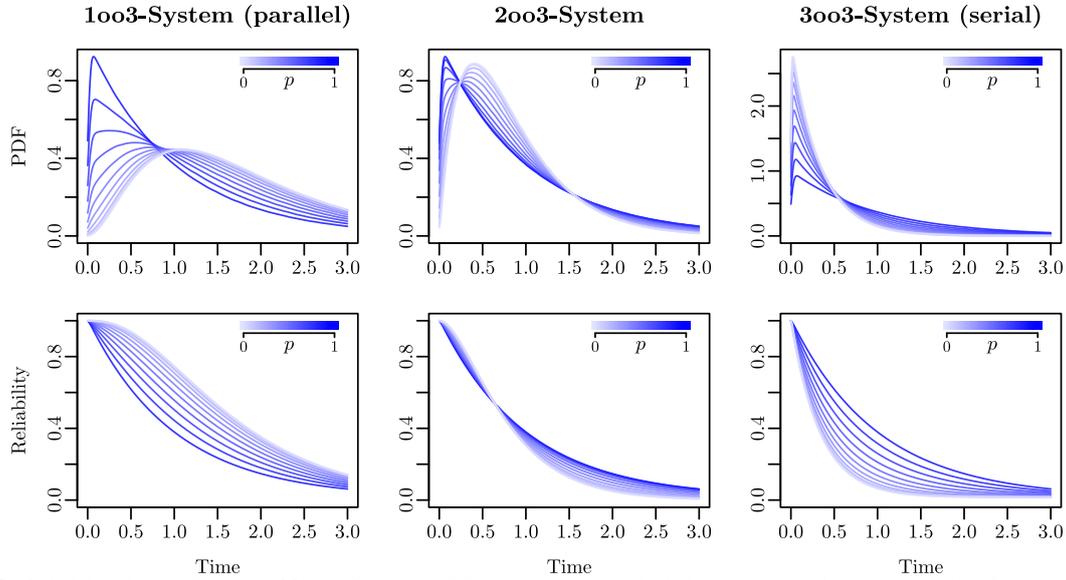

Fig. 5. Probability density function (PDF) of the time to failure (top panels) and reliability function (bottom panels) for the M-out-of-N systems with the marginal Common Cause Failure Model Equation (6) for $p$ ranging from 0 to 1.

In contrast to the previous models, the mean time to failure with the marginal CCF shows nonlinear relationships with $p$ (Figure 6, top panels). The relationships are concave decreasing, sigmoidal and convex increasing for the 1oo3, 2oo3 and 3oo3 systems, respectively.

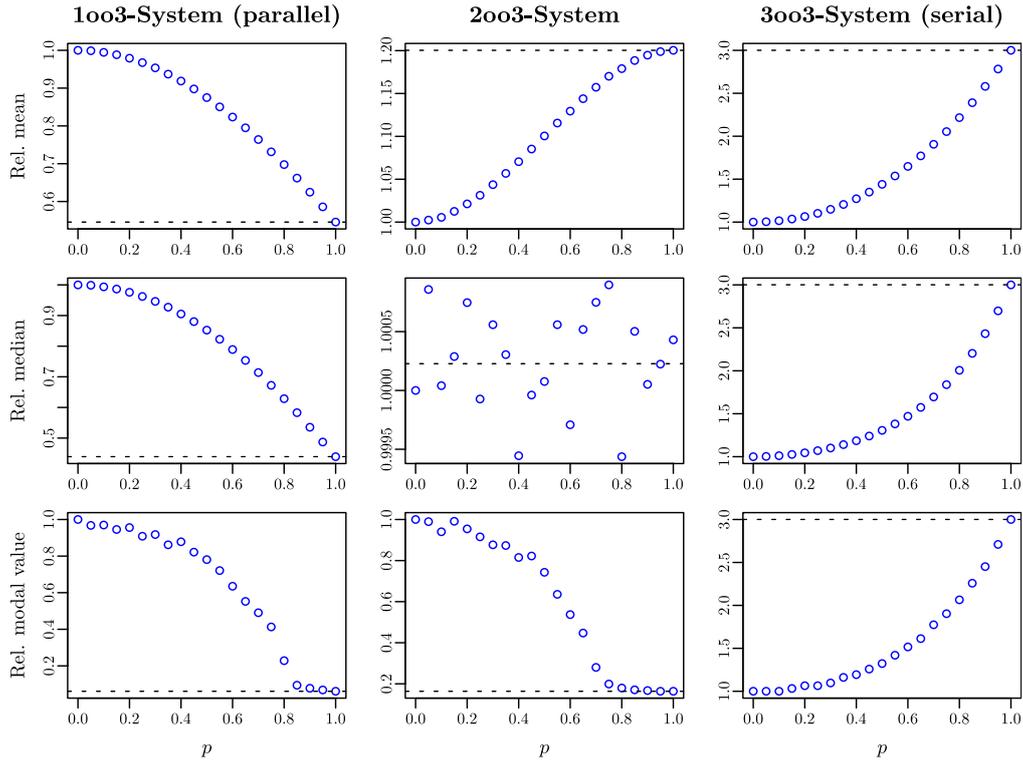

Fig. 6. Relative mean value, median and mode (from the top to the bottom) of the time to failure of MooN systems with the marginal CCF model Equation (6) and $p$ ranging from 0 to 1

## 4. Simulation Tool

The results are obtained by Monte Carlo simulation of the times to failure of the different MooN systems. On average, $10^7$ simulations of the time to failure $T$ given by Equation (3) are executed for each of the three component interdependency models Equations (4), (5) and (6), each of the 1oo3, 2oo3 and 3oo3 architectures, and each value for $p$ ranging from zero to one (20 instances). This corresponds to 1,8 billion random variable draws. The PDF and reliability function in Figures 1, 3, and 5 are obtained using Gaussian kernel density while the means, medians, modes and standard deviations in Figures 2, 4, and 6 are calculated empirically over the simulation samples. The computations are performed in R. The code is given in Figure 7. An online simulation platform is provided, where the parameters can be varied and the code adjusted to obtain customized results (Julitz and Tordeux, 2023).

```r
1   library(Matrix)
2   ## Setting of the parameters M, N and k (shape of Weibull distribution)
        with nb Monte Carlo simulations.
3   N=3; M=2; k=1; nb=1e5
4
5   ## Monte Carlo simulation of the 3 component dependency models
6   sp=seq(0,1,length.out=20)
7   T1=NULL; T2=NULL; T3=NULL
8
9   T=function(p){
10    x0=rweibull(nb,k)
11    Y=(1-p)*matrix(rweibull(nb*N,k),nb,N)+p*matrix(rep(x0,N),nb,N)
12    apply(Y,1,sort)[N-M+1,]}
13
14  for(p in sp){
15    T1=cbind(T1,T(p))
16    p0=runif(nb)<p;T2=cbind(T2,T(matrix(rep(p0,N),nb,N)))
17    T3=cbind(T3,T(matrix(runif(nb*N)<p,nb,N)))}
```

Fig. 7. R code for the Monte Carlo Simulation of M-out-of-N system time to failure Equation (3) with the three component dependency models Equations (4), (5) and (6). Online Simulation Platform (Julitz and Tordeux, 2023).

## 5. Discussion

The analysis of the results of this study provides insights into the complex interaction between component dependencies and the reliability of M-out-of-N (MooN) system architectures. Examination of the three interdependency models - linear dependency, global common causes of failure (GCCF) and marginal common causes of failure (MCCF) - reveals insightful failure behaviors. The **linear dependency** shows different effects on various MooN architectures. While it systematically degrades the reliability of the 1oo3 system, it shows a systematic improvement for the 3oo3 system. The 2oo3 system, on the other hand, shows mixed results, indicating the complexity of the effects of dependencies in different redundancy systems. The **GCCF model** emphasizes the importance of asymmetric distributions in certain redundancy architectures. In particular, it improves the reliability of the 3oo3 system while degrading that of the 1oo3 system. The 2oo3 system shows an intermediate performance. These results point to specific failure mechanisms that should be considered when analyzing redundancy systems. The **MCCF modeling** shows qualitative similarities with the other models. The 2oo3 system continues to show mixed results, with improvements in the modal time to failure, degradation of the mean and no effect on the median. The observed non-linear relationships of mean time to failure with $p$ indicate that the effects of dependence do not scale linearly with the dependence parameter $p$. This could indicate that critical thresholds exist above which reliability is unexpectedly affected. Looking at the different MooN architectures underlines the need to consider dependencies specific to the system configuration. Each architecture - 1oo3, 2oo3 and 3oo3 - reacts differently to interdependencies, with parallel systems being more affected than serial systems.

## 6. Conclusion

This work provides a deeper insight into the impact of component dependencies on the reliability of various M-out-of-N systems. The different responses of the architectures and the non-linear relationships observed underline the complexity of the topic. The findings are particularly important for safety-critical applications such as advanced driver assistance systems (ADAS) for automated driving. Three research questions (RQ) were posed.

**RQ 1**: *How can the dependent failure behaviour of systems be modelled in a minimal way?* Three mathematical models were presented that describe the dependent failure behavior of systems with $k$ components in a compact form (Eqs. 4, 5, 6). Limitation: Due to the general formulation of the models, application to specific systems is difficult to implement.

**RQ 2:** *What generalizations can be made about the impact of dependent redundancy of serial, parallel, and M-out-of-N architectures on system reliability?* The Monte Carlo simulation results of the time to failure of the $MooN$ systems show generic characteristics of the three component dependency models, namely linear, global CCF and marginal CCF models. As expected, the dependency systematically degrades the reliability of the 1oo3 parallel system while it leads to improvements for the 3oo3 serial system. Interestingly, the 2oo3 system shows intermediate results, negative in average but positive for the modal value, when the median remains unaffected (CCF models). Limitation: The analysis was restricted to 1oo3, 2oo3 and 3oo3 systems.

**RQ 3:** *How can the model be used in a user-friendly way?* The models were integrated into an application-oriented online simulation platform based on a Monte Carlo simulation using an R script (Julitz and Tordeux, 2023). The integrated code allows to set the parameters N and M of the MooN systems, distribution parameters and the number of simulations to be set. This allows the analysis of different dependent systems. Limitations: The possibility of analysis using the online tool is limited to simple MooN architectures. If a combination of MooN systems is required for a specific application, the code must be adapted accordingly.

The study focuses on failure density, reliability function and location statistics for the time to failure (mean, median and mode). Other indicators should reveal interesting properties such as failure rate or statistics for the time to failure distribution such as standard deviation, skewness or kurtosis. Further investigations could be done on the modal value, when the median remains unaffected (CCF models. A system with only three components is analysed. Further investigations could be carried out with more complex MooN systems, such as 2oo4 or 3oo4 systems and a combination of them in e.g. serial-parallel and parallel-serial systems to allow more flexibility in the customization of the models. For lack of simplicity, the component times to failure are mixtures of random variables with identical Weibull distributions with a shape parameter equal to one (i.e., exponential distribution). Further research could be carried out using different distributions to account for time-dependent effects. In addition, each component could have a specific parameter setting. A dynamic approach based on stochastic processes could also better describe time-dependent effects, e.g. for modelling standby systems and other complex redundant architectures.